\documentclass{sig-alternate}
\usepackage{fdiaz}
\newcommand{\vocabulary}[0]{\fdset{V}}

\newcommand{\query}[0]{q}

\newcommand{\numTerms}[0]{|\vocabulary|}

\newcommand{\embeddingSize}[0]{k}
\newcommand{\embeddingMatrix}[0]{\fdmat{U}}

\newcommand{\queryTermVector}[0]{\fdvec{\query}}

\newcommand{\word}[0]{w}
\newcommand{\targetWord}[0]{w}
\newcommand{\sourceContext}[0]{c}
\newcommand{\LM}[0]{\theta}
\newcommand{\corpusLM}[0]{\LM_C}
\newcommand{\negContext}[0]{\overline{\targetWord}}

\newcommand{\wordVectorFn}[0]{\phi}
\newcommand{\contextVectorFn}[0]{\psi}
\newcommand{\wordVector}[1]{\wordVectorFn(#1)}
\newcommand{\contextVector}[1]{\contextVectorFn(#1)}

\newcommand{\uniformLoss}{\mathcal{L}}

\newcommand{\windowSize}{m}

\newcommand{\topic}{p_t}
\newcommand{\doc}{d}
\newcommand{\docP}{p_{\doc}}
\newcommand{\docPrime}{\doc'}
\newcommand{\docPrimeP}{p_{\docPrime}}
\newcommand{\queryP}{p_{\query}}
\newcommand{\expandedQueryP}{\queryP^1}
\newcommand{\expansionP}{p_{\query^+}}
\newcommand{\corpusP}{p_c}

\newcommand{\numExpansionTerms}{k}

\makeatletter
\def\@copyrightspace{\relax}
\makeatother

\def\sharedaffiliation{%
\end{tabular}
\begin{tabular}{c}}
\begin{document}
\title{Query Expansion with Locally-Trained Word Embeddings}
 \numberofauthors{3}
 \author{
 \alignauthor Fernando Diaz\\
 \alignauthor Bhaskar Mitra\\
 \alignauthor Nick Craswell\\
 \sharedaffiliation
 \affaddr{Microsoft}\\
 \email{\{fdiaz,bmitra,nickr\}@microsoft.com}
 }
\maketitle
\begin{abstract}
   Continuous space word embeddings have received a great deal of attention in the natural language processing and machine learning communities for their ability to model term similarity and other relationships.  We study the use of term relatedness in the context of query expansion for \emph{ad hoc} information retrieval.  We demonstrate that word embeddings such as word2vec and GloVe, when trained globally, underperform corpus and query specific embeddings for retrieval tasks.  These results suggest that other tasks benefiting from global embeddings may also benefit from local embeddings.  
\end{abstract}

\section{Introduction}
\label{sec:introduction}

Continuous space embeddings such as word2vec \cite{mikolov:word2vec} or GloVe \cite{glove:2014} project terms in a vocabulary to a dense, lower dimensional space.  Recent results in the natural language processing community demonstrate the effectiveness of these methods for analogy and word similarity tasks.  In general, these approaches provide global representations of words; each word has a fixed representation, regardless of any discourse context.  While a global representation provides some advantages, language use can vary dramatically by topic.  For example, ambiguous terms can easily be disambiguated given local information in immediately surrounding words \cite{harris:distributional-structure,yarowsky:one-sense-per-collocation}.  The window-based training of word2vec style algorithms exploits this distributional property.

A global word embedding, even when trained using local windows, risks capturing only coarse representations of those topics dominant in the corpus.  While a particular embedding may be appropriate for a specific word within a sentence-length context globally, it may be entirely inappropriate within a specific topic.  Gale \emph{et al.} refer to this as the `one sense per discourse' property \cite{gale:one-sense-per-discourse}.  Previous work by Yarowsky demonstrates that this property can be successfully combined with information from nearby terms for word sense disambiguation \cite{yarowsky:unsupervised-disambiguation}.  Our work extends  this approach to word2vec-style training in the context word similarity.    

For many tasks that require topic-specific linguistic analysis, we argue that topic-specific representations should outperform global representations.  Indeed, it is difficult to imagine a natural language processing task that would \emph{not} benefit from an understanding of the local topical structure.  Our work focuses on a query expansion, an information retrieval task where we can study different lexical similarity methods with an extrinsic evaluation metric (i.e. retrieval metrics).  Recent work has demonstrated that similarity based on global word embeddings can be used to outperform classic pseudo-relevance feedback techniques \cite{sordoni:qe-embeddings,ALMasri2016}.

We propose that embeddings be learned on topically-constrained corpora, instead of large topically-unconstrained corpora.  In a retrieval scenario, this amounts to retraining an embedding on documents related to the topic of the query.  We present local embeddings which capture the nuances of topic-specific language better than global embeddings.  There is substantial  evidence that global methods underperform local methods for information retrieval tasks such as query expansion \cite{xu:local-global-qe}, latent semantic analysis \cite{hull:local-lsi,schutze:local-routing,singhal:local-routing}, cluster-based retrieval \cite{tombros:querysensitive,tombros:ipm2002,willet:query-specific-clustering}, and term clustering \cite{attar:prf}. We demonstrate that the same holds true when using word embeddings for text retrieval.

\section{Motivation}
\label{sec:motivation}

For the purpose of motivating our approach, we will restrict ourselves to word2vec although other methods behave similarly \cite{levy:word2vec-matrix-factorization}.  These algorithms involve discriminatively training a neural network to predict a word given small set of context words.  More formally, given a target word $\targetWord$ and observed context $\sourceContext$, the instance loss is defined as,
\begin{align*}
   \ell(\targetWord,\sourceContext)&=\log\sigma(\wordVector{\targetWord}\cdot\contextVector{\sourceContext}) \\
   &+ \eta \cdot \mathbb{E}_{\negContext\sim \corpusLM}\left[ \log\sigma(-\wordVector{\targetWord}\cdot\contextVector{\negContext}) \right]
\end{align*}
where $\wordVectorFn : \vocabulary \rightarrow \Re^k$ projects a term into a $k$-dimensional embedding space, $\contextVectorFn : \vocabulary^\windowSize \rightarrow \Re^k$ projects a set of $\windowSize$ terms into a  $k$-dimensional embedding space, and $\negContext$ is a randomly sampled `negative' context.  The parameter $\eta$ controls the sampling of random negative terms.  These matrices are estimated over a set of contexts sampled from a large corpus and minimize the expected loss,
\begin{align}
   \uniformLoss_c&=\mathbb{E}_{\targetWord,\sourceContext\sim \corpusP} \left[\ell(\targetWord,\sourceContext)\right]\label{eq:uniformLoss}
\end{align}
where $\corpusP$ is the distribution of word-context pairs in the training corpus and can be estimated from corpus statistics.  

While using corpus statistics may make sense absent any other information, oftentimes we know that our analysis will be topically constrained.  For example, we might be analyzing the `sports' documents in a collection.  The language in this domain is more specialized and the distribution over word-context pairs is unlikely to be similar to $\corpusP(\targetWord,\sourceContext)$.  In fact, prior work in information retrieval suggests that documents on subtopics in a collection have very different unigram distributions compared to the whole corpus \cite{clarity}.  Let $\topic(\targetWord,\sourceContext )$ be the probability of observing a word-context pair conditioned on the topic $t$.  

The expected loss under this distribution is \cite{shimodaira:covariate-shift},

\begin{align}
\uniformLoss_t &= \mathbb{E}_{\targetWord,\sourceContext\sim \corpusP} \left[\frac{\topic(\targetWord,\sourceContext)}{\corpusP(\targetWord,\sourceContext)}\ell(\targetWord,\sourceContext)\right]\label{eq:topicLoss}
\end{align}
In general, if our corpus consists of sufficiently diverse data (e.g. Wikipedia), the support of $\topic(\targetWord,\sourceContext)$ is much smaller than and contained in that of $\corpusP(\targetWord,\sourceContext)$.  The loss, $\ell$, of a context that occurs more frequently in the topic, will be amplified by the importance weight $\omega=\frac{\topic(\targetWord,\sourceContext)}{\corpusP(\targetWord,\sourceContext)}$.  Because topics require specialized language, this is likely to occur; at the same time, these contexts are likely to be underemphasized in training a model according to Equation \ref{eq:uniformLoss}.  
\begin{figure}
   \centering
   \includegraphics[width=2.5in]{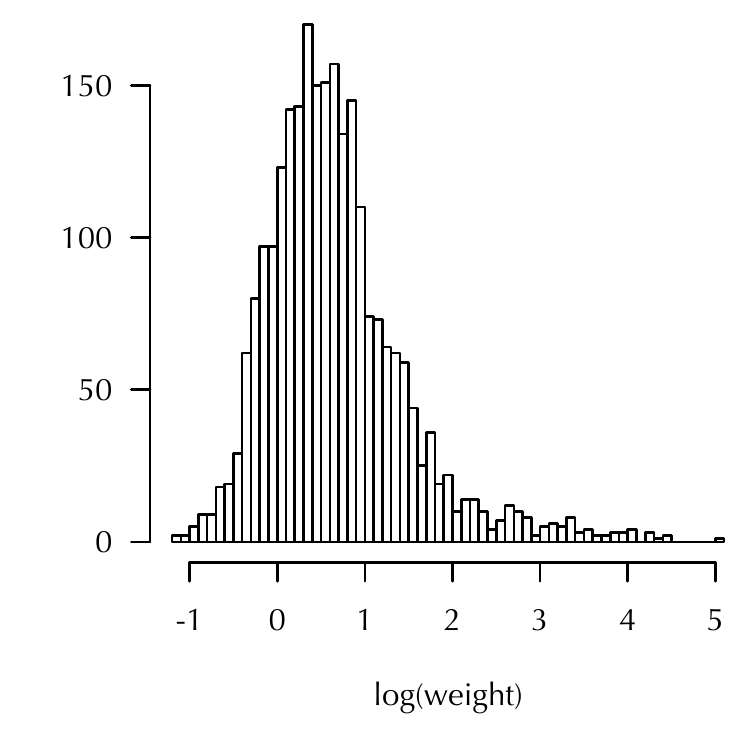}
   \caption{Importance weights for terms occurring in documents related to `argentina pegging dollar' relative to frequency in gigaword.}\label{fig:global-vs-local-importance-weights}
\end{figure}

In order to quantify this, we took a topic from a TREC \emph{ad hoc} retrieval collection (see Section \ref{sec:methods} for details) and computed the importance weight for each term occurring in the set of on-topic documents.  The histogram of weights $\omega$ is presented in Figure~\ref{fig:global-vs-local-importance-weights}.  While larger probabilities are expected since the size of a topic-constrained vocabulary is smaller, there are a non-trivial number of terms with much larger importance weights.  If the loss, $\ell(w)$, of a word2vec embedding is worse for these words with low $\corpusP(w)$, then we expect these errors to be exacerbated for the topic. 

\begin{figure}
   \centering
   \includegraphics[width=2.5in]{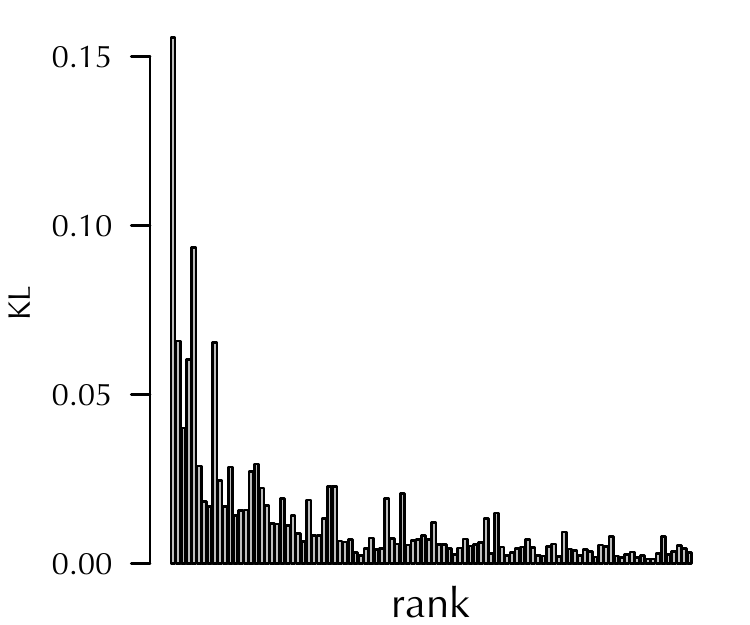}
   \caption{Pointwise Kullback-Leibler divergence for terms occurring in documents related to `argentina pegging dollar' relative to frequency in gigaword.}\label{fig:global-vs-local-kld}
\end{figure}

Of course, these highly weighted terms may have a low value for $\topic(w)$ but a very high value relative to the corpus.  We can adjust the weights by considering the pointwise Kullback-Leibler divergence for each word $w$, 
\begin{align}
   D_w(\topic \| \corpusP) &= \topic(w)\log\frac{\topic(w)}{\corpusP(w)}
\end{align}
Words which have a much higher value of $\topic(w)$ than $\corpusP(w)$ \emph{and} have a high absolute value of $\topic(w)$ will have high pointwise KL divergence.  Figure \ref{fig:global-vs-local-kld} shows the divergences for the top 100 most frequent terms in $\topic(w)$.  The higher ranked terms (i.e. good query expansion candidates)  tend to have much higher probabilities than found in $\corpusP(w)$.  If the loss on those words is large, this may result in poor embeddings for the most important words for the topic.

A dramatic change in distribution between the corpus and the topic has implications for performance precisely because of the objective used by word2vec (i.e. Equation \ref{eq:uniformLoss}).  The training emphasizes word-context pairs occurring with high frequency in the corpus.  We will demonstrate that, even with heuristic downsampling of frequent terms in word2vec, these techniques result in inferior performance for specific topics.  

Thus far, we have sketched out why using the corpus distribution for a specific topic may result in undesirable outcomes.  However,  it is even unclear that $\topic(\targetWord|\sourceContext)=\corpusP(\targetWord|\sourceContext)$.  In fact, we suspect that $\topic(\targetWord|\sourceContext)\neq \corpusP(\targetWord|\sourceContext)$ because of the `one sense per discourse' claim \cite{gale:one-sense-per-discourse}.  We can qualitatively observe the difference in $\corpusP(\targetWord|\sourceContext)$ and $\topic(\targetWord|\sourceContext)$ by training two word2vec models: the first on the large, generic Gigaword corpus and the second on a topically-constrained subset of the gigaword.  We present the most similar terms to `cut' using both a global embedding and a topic-specific embedding in Figure \ref{fig:global-vs-local-terms}.  In this case, the topic is `gasoline tax'.  As we can see, the `tax cut' sense of `cut' is emphasized in the topic-specific embedding.  

\begin{figure}
   \centering
   \begin{tabular}{cc}
      global & local \\
      \hline
      cutting & tax \\
      squeeze & deficit \\
      reduce & vote \\
      slash & budget \\
      reduction & reduction \\
      spend & house \\
      lower & bill \\
      halve & plan \\
      soften & spend \\
      freeze & billion       
      \end{tabular}
   \caption{Terms similar to `cut' for a word2vec model trained on a general news corpus and another trained only on documents related to `gasoline tax'.}\label{fig:global-vs-local-terms}
\end{figure}

\section{Local Word Embeddings}
\label{sec:algorithms}
The previous section described several reasons why a global embedding may result in overgeneral word embeddings.  In order to perform topic-specific training, we need a set of topic-specific documents.  In information retrieval scenarios users rarely provide the system with examples of topic-specific documents, instead providing a small set of keywords.

Fortunately, we can use information retrieval techniques to generate a query-specific set of topical documents.  Specifically, we adopt a language modeling approach to do so \cite{croft:lmbook}.  In this retrieval model, each document is represented as a maximum likelihood language model estimated from document term frequencies.  Query language models are estimated similarly, using term frequency in the query.  A document score then, is the Kullback-Leibler divergence between the query and document language models,
\begin{align}
   D(\queryP \| \docP) & = \sum_{\word\in\vocabulary} \queryP(\word)\log\frac{\queryP(\word)}{\docP(\word)}\label{eq:ql}
\end{align}
Documents whose language models are more similar to the query language model will have a \emph{lower} KL divergence score.  For consistency with prior work, we will refer to this as the query likelihood score of a document.  

The scores in Equation \ref{eq:ql} can be passed through a softmax function to derive a multinomial over the entire corpus  \cite{victor:RM},
\begin{align}
   p(\doc) &= \frac{\exp(-D(\queryP \| \docP))}{\sum_{\docPrime}\exp(-D(\queryP \| \docPrimeP))}
\end{align}
Recall in Section \ref{sec:motivation} that training a word2vec model weights word-context pairs according to the corpus frequency.  Our query-based multinomial, $p(\doc)$, provides a weighting function capturing the documents relevant to this topic.  Although an estimation of the topic-specific documents from a query will be imprecise (i.e. some nonrelevant documents will be scored highly), the language use tends to be consistent with that found in the known relevant documents. 

We can train a local word embedding using an arbitrary optimization method by sampling documents from $p(\doc)$ instead of uniformly from the corpus.  In this work, we use word2vec, although any method that operates on a sample of documents can be used.

\section{Query Expansion with Word Embeddings}
When  using language models for retrieval, query expansion involves estimating an alternative to $\queryP$.  Specifically, when each expansion term is associated with a weight, we normalize these weights to derive the expansion language model, $\expansionP$.  This language model is then interpolated with the original query model,
\begin{align}
   \expandedQueryP(\word) &= \lambda \queryP(\word)+  (1-\lambda) \expansionP(\word)
\end{align}
This interpolated language model can then be used with Equation \ref{eq:ql} to rank documents \cite{umass-hard2004}.  We will refer to this as the expanded query score of a document.

Now we turn to using word embeddings for query expansion.  Let $\embeddingMatrix$ be an $\numTerms\times\embeddingSize$ term embedding matrix.  If $\queryTermVector$ is a $\numTerms\times 1$ column term vector for a query, then the expansion term weights are $\embeddingMatrix\fdtrans{\embeddingMatrix}\queryTermVector$.  We then take the top $\numExpansionTerms$ terms, normalize their weights, and compute $\expansionP(\word)$.

We consider the following alternatives for $\embeddingMatrix$.  The first approach is to use a global model trained by sampling documents uniformly.   The second approach, which we propose in this paper, is to use a local model trained by sampling documents from $p(d)$.

\section{Methods}
\label{sec:methods}
\subsection{Data}
To evaluate the different retrieval strategies described in Section \ref{sec:algorithms}, we use the following datasets.  Two newswire datasets, trec12 and robust, consist of the newswire documents and associated queries from TREC \emph{ad hoc} retrieval evaluations. The trec12 corpus consists of Tipster disks 1 and 2; and the robust corpus consists of Tipster disks 4 and 5.  Our third dataset, web, consists of the ClueWeb 2009 Category B Web corpus. For the Web corpus, we only retain documents with a Waterloo spam rank above 70.\footnote{\url{https://plg.uwaterloo.ca/~gvcormac/clueweb09spam/}}  We present corpus statistics in Table \ref{tab:corpus-statistics}.

We consider several publicly available global embeddings.  We use four GloVe embeddings of different dimensionality trained on the union of Wikipedia and Gigaword documents.\footnote{\url{http://nlp.stanford.edu/data/glove.6B.zip}}  We use one publicly available word2vec embedding trained on Google News documents.\footnote{\url{https://code.google.com/archive/p/word2vec/}}   We also trained a global embedding for trec12 and robust using the entire corpus.  Instead of training a global embedding on the large web collection, we use a GloVe embedding trained on Common Crawl data.\footnote{\url{http://nlp.stanford.edu/data/glove.840B.300d.zip}}  

\begin{table}
\caption{Corpora used for retrieval and local embedding training.}\label{tab:corpus-statistics}
	\centering
\begin{tabular}{lrrc}
   & \multicolumn{1}{c}{docs} & \multicolumn{1}{c}{words} & queries \\
   \hline
   trec12 & 469,949 & 438,338 & 150 \\
   robust & 528,155 & 665,128 & 250 \\
   web    & 50,220,423 & 90,411,624 & 200 \\
   news   & 9,875,524 & 2,645,367 & - \\
   wiki   & 3,225,743 & 4,726,862 & - \\
\end{tabular}
\end{table}

We train local embeddings with word2vec using one of three retrieval sources.  First, we consider documents retrieved from the target corpus of the query (i.e. trec12, robust, or web).  We also consider training a local embedding by performing a retrieval on large auxiliary corpora.  We use the Gigaword corpus as a large auxiliary news corpus.  We hypothesize that retrieving from a larger news corpus will provide substantially more local training data than a target retrieval.  We also use a Wikipedia snapshot from December 2014.  We hypothesize that retrieving from a large, high fidelity corpus will provide cleaner language than that found in lower fidelity target domains such as the web.  Table \ref{tab:corpus-statistics} shows the relative magnitude of these auxiliary corpora compared to the target corpora. 

All corpora in Table \ref{tab:corpus-statistics} were stopped using the SMART stopword list\footnote{\url{http://jmlr.csail.mit.edu/papers/volume5/lewis04a/a11-smart-stop-list/english.stop}} and stemmed using the Krovetz algorithm \cite{kstem}.  We used the Indri implementation for indexing and retrieval.\footnote{\url{http://www.lemurproject.org/indri/}}

\subsection{Evaluation}
We consider several standard retrieval evaluation metrics, including NDCG@10 and interpolated precision at standard recall points \cite{NDCG,vanrijsbergen:IR}.  NDCG@10 provides insight into performance specifically at higher ranks.  An interpolated precision recall graph  describes system performance throughout the entire ranked list.

\subsection{Training}
All retrieval experiments were conducted by performing 10-fold cross-validation across queries.  Specifically, we cross-validate the number of expansion terms, $\numExpansionTerms\in[5-500]$, and interpolation weight, $\lambda \in[0,1]$.  For local word2vec training, we cross-validate the  learning rate $\alpha\in\{0.1,0.01,0.001\}$.

All word2vec training used the publicly available word2vec cbow implementation.\footnote{\url{https://code.google.com/p/word2vec/}} When training the local models, we sampled 1000 documents from $p(\doc)$ with replacement.  To compensate for the much smaller corpus size, we ran word2vec training for 80 iterations.  Local word2vec models use a fixed embedding dimension of 400 although other choices did not significantly affect our results.  Unless otherwise noted, default parameter settings were used.

In our experiments, expanded queries rescore the top 1000 documents from an initial query likelihood retrieval.   Previous results have demonstrated that this approach results in performance nearly identical with an expanded retrieval at a much lower cost \cite{diaz:crm}.  Because publicly available embeddings may have tokenization inconsistent with our target corpora, we restricted the vocabulary of candidate expansion terms to those occurring in the initial retrieval.  If a candidate term was not found in the vocabulary of the embedding matrix, we searched for the candidate in a stemmed version of the embedding vocabulary.  In the event that the candidate term was still not found after this process, we removed it from consideration.

\section{Results}
\label{sec:results}
We present results for retrieval experiments in Table \ref{tab:results}.  We find that embedding-based query expansion outperforms our query likelihood baseline across all conditions.  When using the global embedding, the news corpora benefit from the various embeddings in different situations.  Interestingly, for trec12, using an embedding trained on the target corpus significantly outperforms all other global embeddings, despite using substantially less data to estimate the model.  While this performance may be due to the embedding having a tokenization consistent with the target corpus, it may also come from the fact that the corpus is more representative of the target documents than other embeddings which rely on online news or are mixed with non-news content.  To some extent this supports our desire to move training closer to the target distribution.  
\begin{table*}
   \centering
\caption{Retrieval results comparing query expansion based on various global and local embeddings.  Bolded numbers indicate the best expansion in that class of embeddings. Wilcoxon signed rank test between bolded numbers indicates statistically significant improvements ($p<0.05$) for all collections.}\label{tab:results}
   
   \begin{tabular}{lc|cccccc|ccccc}
    &    & \multicolumn{6}{c|}{global} & \multicolumn{3}{c}{local}\\
    &    & \multicolumn{4}{c}{wiki+giga} & gnews & target & target & giga & wiki\\
    & QL & 50 & 100 & 200 & 300 & 300 & 400 & 400 & 400 & 400 \\
    \hline
    trec12& 0.514 & 0.518 & 0.518 & 0.530 & 0.531 & 0.530 & \textbf{0.545} & 0.535 & \textbf{0.563}\textsuperscript{*} & 0.523\\
    robust& 0.467 & 0.470 & 0.463 & 0.469 & 0.468 & \textbf{0.472} & 0.465 & 0.475 & \textbf{0.517}\textsuperscript{*} & 0.476\\
    web& 0.216 & 0.227 & 0.229 & 0.230 & \textbf{0.232} & 0.218 & 0.216 & 0.234 & 0.236 & \textbf{0.258}\textsuperscript{*}
   \end{tabular}

\end{table*}

\begin{figure*}
   \centering
   \includegraphics[width=6.5in]{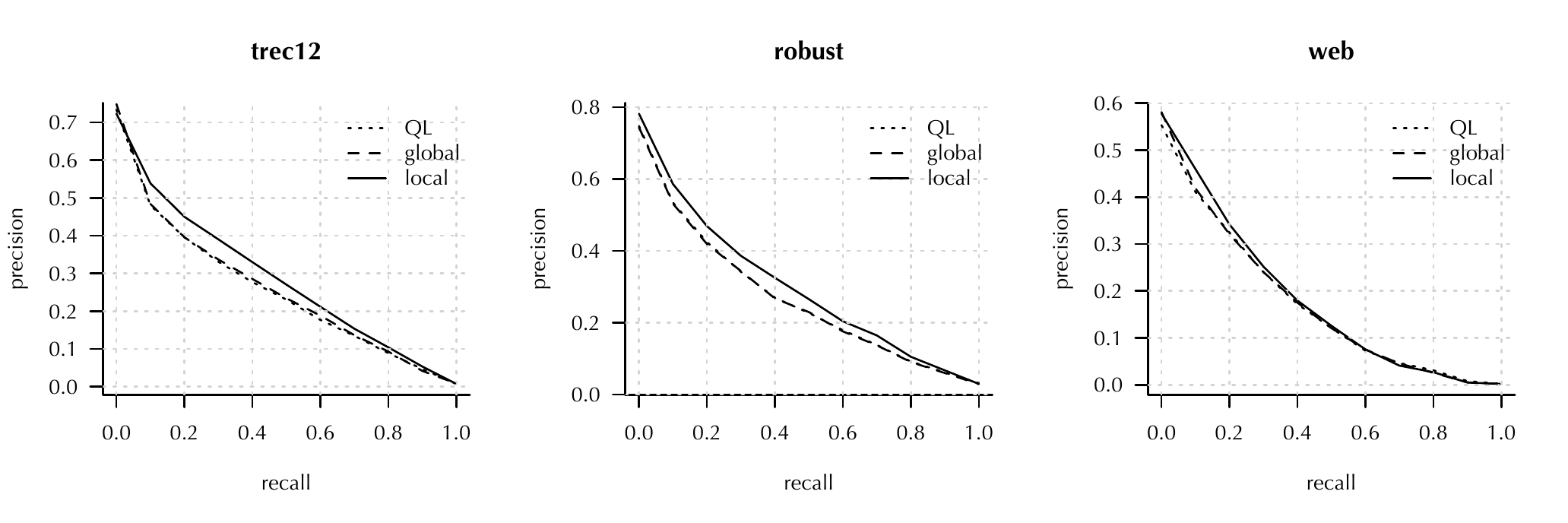}
\caption{Interpolated precision-recall curves for query likelihood, the best global embedding, and the best local embedding from Table \ref{tab:results}.  }\label{fig:results:interpolated-pr}
\end{figure*}

Across all conditions, local embeddings significantly outperform global embeddings for query expansion.  For our two news collections, estimating the local model using a retrieval from the larger Gigaword corpus led to substantial improvements.  This effect is almost certainly due to the Gigaword corpus being similar in writing style to the target corpus but, at the same time, providing significantly more relevant content \cite{diaz:ee}.  As a result, the local embedding is trained using a larger variety of topical material than if it were to use a retrieval from the smaller target corpus.  An embedding trained with a retrieval from Wikipedia tended to perform worse most likely because the language is dissimilar from news content.  Our web collection, on the other hand, benefitted more from embeddings trained using retrievals from the general Wikipedia corpus.  The Gigaword corpus was less useful here because news-style language is almost certainly not representative of general web documents.

Figure \ref{fig:results:interpolated-pr} presents interpolated precision-recall curves comparing the baseline, the best global query expansion method, and the best local query expansion method.  Interestingly, although global methods achieve strong performance for NDCG@10, these improvements over the baseline are not reflected in our precision-recall curves.  Local methods, on the other hand, almost always strictly dominate both the baseline and global expansion across all recall levels.

The results support the hypothesis that local embeddings provide better similarity measures than global embeddings for query expansion. In order to understand why, we first compare the performance differences between local and global embeddings. Figure \ref{fig:global-vs-local-kld} suggests that we should adopt a local embedding when the local unigram language model deviates from the corpus language model.  To test this, we computed the KL divergence between the local unigram distribution, $\sum_d p(w|d)p(d)$, and the corpus unigram language model \cite{clarity}.  We hypothesize that, when this value is high, the topic language is different from the corpus language and the global embedding will be inferior to the local embedding.  We tested the rank correlation between this KL divergence and the relative performance of the local embedding with respect to the global embedding.  These correlations are presented in Table \ref{tab:clarity}.  Unfortunately, we find that the correlation is  low, although it is positive across collections.

\begin{table}
   \caption{Kendall's $\tau$ and Spearman's $\rho$ between improvement in NDCG@10 and local KL divergence with the corpus language model.  The improvement is measured for the best local embedding over the best global embedding.}\label{tab:clarity}
   \centering
   \begin{tabular}{lcc}
      & $\tau$ & $\rho$\\
      \hline
      trec12 & 0.0585 & 0.0798\\
      robust & 0.0545 & 0.0792\\
      web    & 0.0204 & 0.0283
   \end{tabular}
\end{table}

\begin{figure}[t]
   \includegraphics[width=3in]{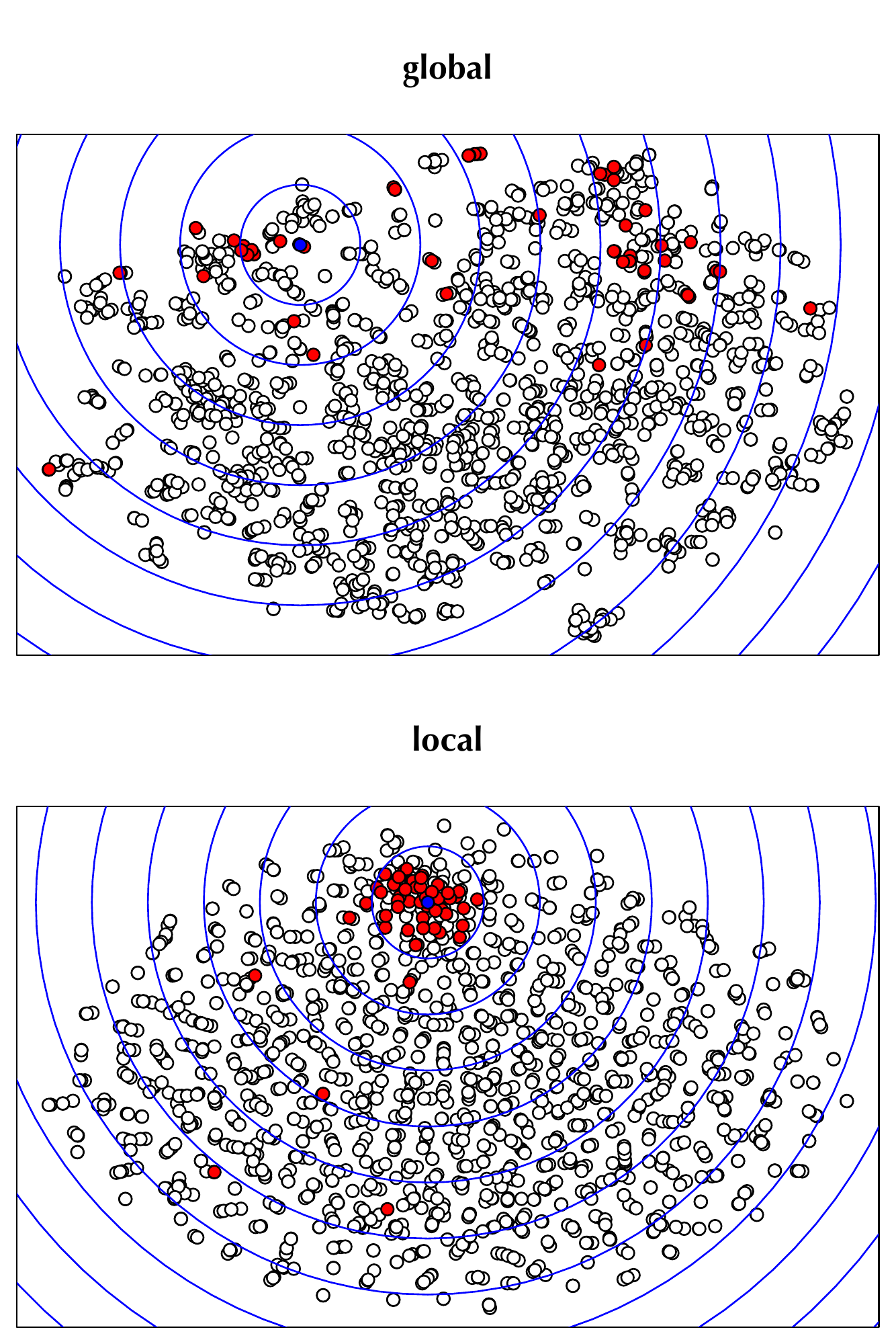}
   \caption{Global versus local embedding of highly relevant terms.  Each point represents a candidate expansion term.  Red points have high frequency in the relevant set of documents.  White points have low or no frequency in the relevant set of documents.  The blue point represents the query.  Contours indicate distance from the query.  }\label{fig:tsne}
\end{figure}

We can also qualitatively analyze the differences in the behavior of the embeddings.  If we have access to the set of documents labeled relevant to a query, then we can compute the frequency of terms in this set and consider those terms with high frequency (after stopping and stemming) to be good query expansion candidates.  We can then visualize where these terms lie in the global and local embeddings.  In Figure \ref{fig:tsne}, we present a two-dimensional projection \cite{tsne} of terms for the query `ocean remote sensing', with those  good candidates highlighted.  Our projection includes the top 50 candidates by frequency and a sample of terms occurring in the query likelihood retrieval.  We notice that, in the global embedding, the good candidates are spread out amongst poorer candidates.  By contrast, the local embedding clusters the candidates in general but also situates them closely around the query.  As a result, we suspect that the similar terms extracted from the local embedding are more likely to include these good candidates.

\section{Discussion}
\label{sec:discussion}

The success of local embeddings on this task should alarm natural language processing researchers using global embeddings as a representational tool.  For one, the approach of learning from vast amounts of data is only effective if the data is appropriate for the task at hand.  And, when provided, much smaller high-quality data can provide much better performance.  Beyond this, our results suggest that the approach of estimating global representations, while computationally convenient, may overlook insights possible at query time, or evaluation time in general.  A similar local embedding approach can be adopted for any natural language processing task where topical locality is expected and can be estimated.  Although we used a query to re-weight the corpus in our experiments, we could just as easily use alternative contextual information (e.g. a sentence, paragraph, or document) in other tasks.  

Despite these strong results, we believe that there are still some open questions in this work.  First, although local embeddings provide effectiveness gains, they can be quite inefficient compared to global embeddings.  We believe that there is opportunity to improve the efficiency by considering offline computation of local embeddings at a coarser level than queries but more specialized than the corpus.  If the retrieval algorithm is able to select the appropriate embedding at query time, we can avoid training the local embedding.  Second, although our supporting experiments (Table \ref{tab:clarity}, Figure \ref{fig:tsne}) add some insight into our intuition, the results are not strong enough to provide a solid explanation.  Further theoretical and empirical analysis is necessary.

\section{Related Work}

\paragraph*{Topical adaptation of models}
The shortcomings of learning a single global vector representation, especially for polysemic words, have been pointed out before \cite{reisinger2010multi}. The problem can be addressed by training a global model with multiple vector embeddings per word \cite{reisinger2010mixture,huang2012improving} or topic-specific embeddings \cite{AAAI159314}. The number of senses for each word may be fixed \cite{neelakantan2015efficient}, or determined using class labels \cite{trask2015sense2vec}. However, to the best of our knowledge, this is the first time that training topic-specific word embeddings has been explored.

Several methods exist in the language modeling community for topic-dependent adaptation of language models \cite{bellegarda2004statistical}. These can lead to performance improvements in tasks such as machine translation \cite{zhao:lm-adaptation} and speech recognition \cite{nanjo2004language}. Topic-specific data may be gathered in advance, by identifying corpus of topic-specific documents. It may also be gathered during the discourse, using multiple hypotheses from N-best lists as a source of topic-specific language. Then a topic-specific language model is trained (or the global model is adapted) online using the topic-specific training data.  A topic-dependent model may be combined with the global model using linear interpolation \cite{iyer:lm-adaptation} or other more sophisticated approaches \cite{federico1996bayesian,kuhn1990cache}. Similarly to the adaptation work, we use topic-specific documents to train a topic-specific model. In our case the documents come from a first round of retrieval for the user's current query, and the word embedding model is trained based on sentences from the topic-specific document set. Unlike the past work, we do not focus on interpolating the local and global models, although this is a promising area for future work. In the current study we focus on a direct comparison between the local-only and global-only approach, for improving retrieval performance.

\paragraph*{Word embeddings for IR}

Information Retrieval has a long history of learning representations of words that are low-dimensional dense vectors. These approaches can be broadly classified into two families based on whether they are learnt based on a \emph{term-document} matrix or \emph{term co-occurence} data. Using the \emph{term-document} matrix for embedding leads to several well-studied approaches such as LSA \cite{deerwester:lsa}, PLSA \cite{hofmann:plsa}, and LDA \cite{blei:lda,wei:lda-retrieval}. The performance of these models varies depending on the task, for example they are known to perform poorly for retrieval tasks unless combined with lexical features \cite{atreya2011latent}. \emph{Term-cooccurence} based embeddings, such as word2vec \cite{mikolov:word2vec,mikolov2013efficient} and \cite{pennington2014glove}, have recently been remarkably popular for many natural language processing and logical reasoning tasks. However, there are relatively less known successful applications of these models in IR. Ganguly \emph{et. al.} \cite{ganguly:word2vec-lm} used the word similarity in the word2vec embedding space as a way to estimate term transformation probabilities in a language modelling setting for retrieval. More recently, Nalisnick \emph{et. al.} \cite{nalisnick2016improving} proposed to model document about-ness by computing the similarity between all pairs of query and document terms using dual embedding spaces. Both these approaches estimate the semantic relatedness between two terms as the cosine distance between them in the embedding space(s). We adopt a similar notion of term relatedness but focus on demonstrating improved retrieval performance using locally trained embeddings.

\paragraph*{Local latent semantic analysis}

Despite the mathematical appeal of latent semantic analysis, several experiments suggest that its empirical performance may be no better than that of ranking using standard term vectors \cite{deerwester:lsa,dumais:trec3,atreya:lsi-sad}.  In order to address the coarseness of corpus-level latent semantic analysis, Hull proposed restricting analysis to the documents relevant to a query \cite{hull:local-lsi}.  This approach significantly improved over corpus-level  analysis for routing tasks, a result that has been reproduced in consequent research \cite{schutze:local-routing,singhal:local-routing}.  Our work can be seen as an extension of these results to more recent techniques such as word2vec.  

\section{Conclusion}
\label{sec:conclusion}
We have demonstrated a simple and effective method for performing query expansion with word embeddings.  Importantly, our results highlight the value of locally-training word embeddings in a query-specific manner.  The strength of these results suggests that other research adopting global embedding vectors should consider local embeddings as a potentially superior representation.  
Instead of using a ``Sriracha sauce of deep learning,'' as embedding techniques like word2vec have been called, we contend that the situation sometimes requires, say, that we make a b\'{e}chamel or a mole verde or a sambal---or otherwise learn to cook.

\bibliographystyle{abbrv}

\bibliography{acl2016-fdiaz,acl2016-bmitra}
\balancecolumns
\end{document}